\documentclass[aps,twocolumn,nofootinbib,showpacs,prd,aps,10pt]{revtex4}
\usepackage[dvips]{graphicx}
\usepackage[english]{babel}
\usepackage[T1]{fontenc}
\usepackage{mathrsfs}
\usepackage[tbtags]{amsmath}
\usepackage{amssymb}
\usepackage{amsxtra}
\usepackage{amsopn}
\usepackage{latexsym}
\usepackage[mathcal]{eucal}
\usepackage{mathtools}

\newcommand{\BE}{\begin{equation}}
\newcommand{\EE}{\end{equation}}
\newcommand{\BA}{\begin{align}}
\newcommand{\EA}{\end{align}}

\begin{document}

\title{Quasigluon lifetime and confinement from first principles}

\author{Fabio Siringo}

\affiliation{Dipartimento di Fisica e Astronomia 
dell'Universit\`a di Catania,\\ 
INFN Sezione di Catania,
Via S.Sofia 64, I-95123 Catania, Italy}

\date{\today}

\begin{abstract}
The mass and the lifetime of a gluon are evaluated from
first principles at finite temperature across the deconfinement 
transition of pure SU(3) Yang-Mills  theory,
by a direct calculation of the pole
of the propagator in the complex plane, using the finite
temperature extension of a massive expansion in the Landau gauge.
Even at $T=0$ the quasigluon lifetime is finite and the gluon
is canceled from the asymptotic states, yielding a microscopic
proof of confinement from first principles. Above the transition
the damping rate is a linear increasing function of temperature as
predicted by standard perturbation theory.

\end{abstract}

\pacs{12.38.Aw, 12.38.Bx, 12.38.Lg, 14.70.Dj}



\maketitle

\maketitle

\section{introduction}
Quarks and gluons are believed to be confined because no free-particle asymptotic states
have ever been observed. However, despite the success of QCD,  no formal proof of confinement
has been derived from first principles yet. The analytical results of standard perturbation theory break down 
in the infrared (IR) where most of our knowledge of non-Abelian theories relies on numerical nonperturbative methods.

In the last decades, important achievements have been reached by simulation of huge lattices and by improved
truncation schemes of Dyson-Schwinger equations (DSE). In the pure gauge sector, the gluon propagator has attracted
a lot of interest because of the direct dynamical information that could be extracted by its detailed knowledge.
Unfortunately, lattice simulations and numerical solutions of DSE have provided a very accurate description of the
propagator in the Euclidean space, but few direct information on the analytic properties that determine the physical
dynamical behavior of the gluon. Actually, the analytic continuation of a limited set of data points to Minkowski 
space is an ill-defined problem. While some evidence for positivity violation has emerged, the numerical attempts 
only give qualitative results at best\cite{dudal14}.
On the other hand, for the study of the hot matter created in heavy ion collisions, 
quasiparticle models are quite successful and make use of temperature-dependent phenomenological
masses and widths for the quasigluons\cite{werner2016,castorina2012,alba2012,greco2011}.

In this paper, for the first time, the real and imaginary part of the gluon mass are evaluated from first
principles across the deconfinement transition of pure SU(3) Yang-Mills  theory, by a direct calculation of the
pole of the gluon propagator in the complex plane as a function of temperature. 
The result is achieved by a finite-temperature extension
of a massive expansion\cite{ptqcd,ptqcd2,analyt,scaling} that was shown to provide very accurate 
analytical expressions for the propagator and
the self energy in the IR. Even in the limit $T\to 0$, the imaginary part of the mass saturates at a finite
value $\gamma\approx 0.48$ GeV, yielding a very short finite lifetime $\tau=1/\gamma$ that eliminates the gluon
from the asymptotic states. Thus the gluon is confined and the quasigluon can only exist as short-lived 
intermediate state at the origin of a gluon-jet event\cite{stingl}.

From a formal point of view, the massive expansion of Ref.\cite{ptqcd2} is obtained by perturbation theory expanding
around a massive zeroth order free propagator in the Landau gauge. As first pointed out in  Ref.\cite{journey} and 
fully discussed in Ref.\cite{comitini}, 
the expansion can
be derived by a variational argument as an expansion around the best vacuum that minimizes the Gaussian effective 
potential\cite{stevenson,var,light,bubble,su2,LR,HT,stancu2,stancu,superc1,superc2,AF}.
A massive vacuum for the gluon is shown to be energetically favoured, but the actual mass scale cannot be 
derived by that method because of the lack of any scale in  Yang-Mills theory. Moreover, the expansion can be further
optimized by a best choice of the subtraction point and can be classified as a special case of 
renormalization-scheme (RS) optimized perturbation theory (OPT)\cite{stevensonRS}.

The massive expansion has many merits. It is based on the Faddeev-Popov gauge-fixed Yang-Mills Lagrangian in the Landau
gauge, namely the same Lagrangian used in most of the lattice simulations. There are no Landau poles in the IR nor
diverging mass terms. No spurious parameters or mass 
counterterms are required, yielding an analytical calculational method from first principles.
Minkowski space is the native environment of the expansion, and Wick rotation is only used as a standard and rigorous
tool for the actual evaluation of the elementary integrals by dimensional regularization. At one loop and $T=0$ the
self energy $\Sigma (p^2)$ and the propagator $\Delta (p^2)$ can be evaluated analytically, 
providing analytic functions that can be easily continued to the Euclidean space where the agreement with the 
lattice data is impressive\cite{scaling}.

The method is very predictive since the adimensional ratio $\sigma(p^2)=\Sigma(p^2)/p^2$
is determined up to an additive constant which, as usual, depends on the renormalization scheme and should be absorbed
by a change of the wave-function renormalization constant. In other words, the derivative of the function 
$\sigma$ is fully
determined yielding a universal prediction for the derivatives of the inverse dressing functions, in
perfect agreement with the lattice data\cite{scaling}, without having to fix any parameter.

Back to the propagator, once the mass scale is fixed by a comparison with the lattice or with the phenomenology,
the only free parameter is the subtraction point. Its change should be absorbed by a change of the wave-function
renormalization constant but it is not, because of the one-loop approximation. 
The residual dependence
of the propagator on the subtraction point, i.e. on the additive constant of the adimensional ratio $\sigma$,
can be further optimized by RD-OPT. Surprisingly, an optimal choice of the additive constant exists that makes the 
neglected higher order terms vanishing, at least in the Euclidean space where a comparison with the lattice data can
be made. Strictly speaking, that only tells us that higher order terms can be written as 
$\Sigma\approx {\rm const.}\times p^2$ and can be absorbed by a change of the subtraction point.
Thus, the optimized one-loop approximation provides reliable analytical functions for the gluon and ghost 
propagator and can be easily continued and studied in the whole complex plane\cite{analyt}.

An important prediction of the calculation is the existence of complex conjugate poles in the gluon propagator.
Their existence was conjectured but not proven before. While usual dispersion relations do not hold in the presence
of complex poles\cite{dispersion}, no violation of causality and unitarity emerges by a careful analysis, as
fully discussed by Stingl more than twenty years ago\cite{stingl}. The imaginary part of the mass leads to
short-lived intermediate states that cannot be present among the asymptotic states. Thus the existence of
complex poles is a direct microscopic proof of confinement.
Moreover, even if the propagator is a gauge dependent function, its poles are gauge-invariant physical
observables\cite{kobes90} and their dependence on temperature would be of primary importance for a microscopic description of
the deconfinement transition.

The extension to finite temperature is straightforward and only requires the evaluation of the
thermal parts of the graphs that are retained in the expansion. Explicit expressions have been derived in the very
different approach of Ref.\cite{serreau} that shares some of
the massive one-loop graphs. Some new crossed graphs are required in the present massive expansion and can be obtained
by a simple derivative. All thermal parts are finite but require a numerical one-dimensional integration.
While the details of the explicit calculation will be published elsewhere, in this paper the trajectory of the poles
of the gluon propagator is studied in the complex plane,  as a function of temperature, in the long wavelength limit.
A crossover is found from a low temperature intrinsically confined gluon to a high temperature thermal quasiparticle.

The non-monotonic behavior of the mass and the linear increase at high temperature are in qualitative agreement with
the predictions of phenomenological quasiparticle models\cite{werner2016,castorina2012,alba2012,greco2011} and
of high temperature perturbative calculations\cite{braaten90}, giving us
more confidence in the genuine physical nature of the poles even at $T=0$.

The paper is organized as follows:
In Section II, some features of the massive expansion at $T=0$ are clarified and discussed together 
with the physical meaning of the poles and their relevance for a microscopic description of
the deconfinement transition; in Section III the extension to finite temperature is described and the trajectory of the
poles is studied in the complex plane, as a function of temperature, in the long wavelength limit where they give the mass
of the quasigluon; In Section IV a general qualitative discussion of deconfinement is given at the light of the present findings.

\section{Complex poles and confinement}
 
\begin{figure}[b] 
\centering
\vskip 20pt
\includegraphics[width=0.12\textwidth,angle=-90]{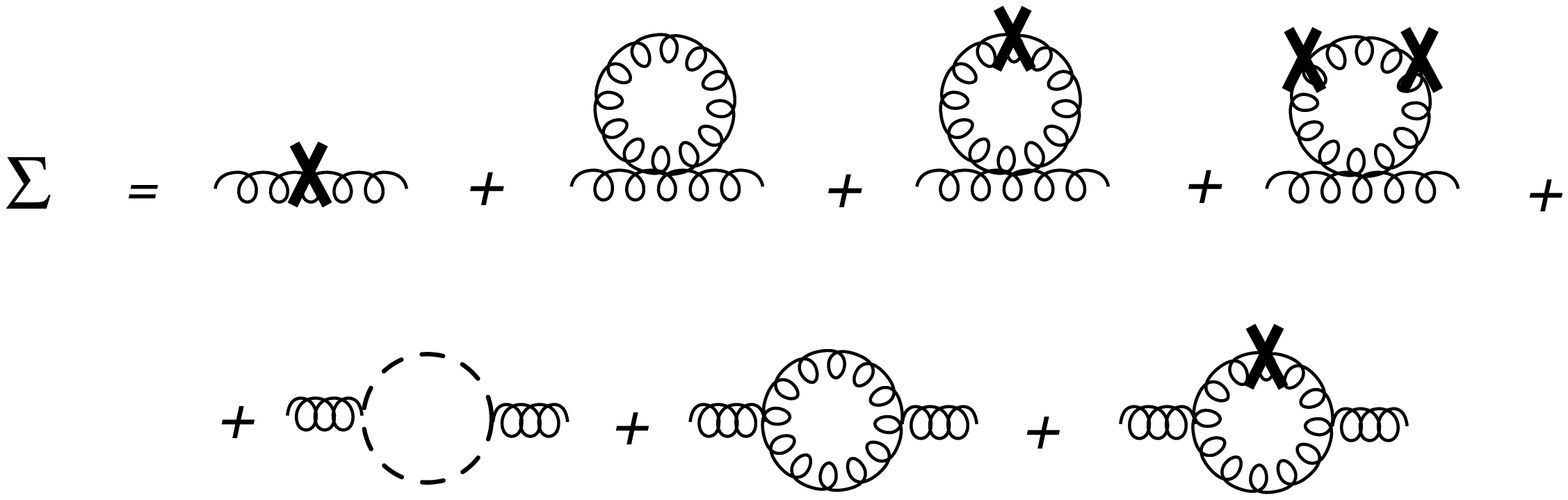}
\caption{Two-point graphs with no more than three vertices and no more than one loop. 
The cross is the counterterm $\delta \Gamma=m_0^2$.}
\label{F1}
\end{figure}

At $T=0$ the gluon propagator of pure SU(3) Yang-Mills theory 
can be evaluated by the optimized one-loop massive expansion of Ref.\cite{ptqcd2}
and the explicit analytical result can be continued to the whole complex plane as 
discussed in Ref.\cite{analyt}.
We refer to those papers for the details of the calculation. 

The massive expansion is obtained by just adding a mass term $m_0^2$ to the quadratic part of the standard
Faddeev-Popov Lagrangian in the Landau gauge and subtracting the same mass term in the interaction 
by a counterterm $\delta\Gamma=m_0^2$. Thus, the total Lagrangian is unchanged and its exact study would
lead to the same results of lattice simulations.

The self energy is expanded by standard perturbation theory
retaining only two-point graphs with no more than three vertices and no more than one loop, as shown in Fig.~\ref{F1}.
The internal gluon lines in the graphs are given by the massive zeroth order propagator
\BE
\Delta _m(p^2) =[-p^2 + m_0^2]^{-1}
\label{Dm}
\EE
and the mass $m_0$ is the only energy scale in the calculation. 
It can only be fixed by comparison with the phenomenology or lattice data.
The counterterm cancels all spurious mass divergences and the expansion can be renormalized by standard
wave function renormalization in a dimensional regularization scheme. At one-loop the graphs can be evaluated
analytically and explicit expressions were reported by Tissier and Wschebor\cite{tissier10,tissier11} 
for most of the graphs in Fig.~\ref{F1}. The crossed graphs in Fig.~\ref{F1}, containing one or more insertions of the counterterm,
are obtained by a simple derivative of the other graphs and explicit expressions can be found in Ref.\cite{ptqcd2}.
We observe that at tree-level, the first graph in Fig.~\ref{F1} cancels the mass shift $m_0^2$ in the propagator and
the renormalized dressed gluon propagator $\Delta(p)$ can be written as
\BE
\Delta(p)=\frac{Z}{-p^2-\Sigma_L(p)}=\frac{J(p)}{-p^2}
\label{G}
\EE
where $Z$ is the wave-function renormalization constant, $J(p)$ is the  dressing function and
$\Sigma_L$ is the sum of all self-energy graphs containing loops. Thus the dynamical generation of mass arises
from loops and no mass would be predicted in QED by the same method.

As usual, at one loop, we can write $Z$ as the product of a finite renormalization constant $z$ times
a diverging factor $1+\alpha\>\delta Z$, where $\alpha$ is some coupling here taken as
$\alpha=3N\alpha_s/(4\pi)$ where $\alpha_s=g^2/(4\pi)$ is the strong coupling.
The dressing function reads
\BE
z\>J(p)^{-1}=1+\alpha\left[F(p^2/m_0^2)-\delta Z\right]
\label{chi}
\EE
where the adimensional function $F(s)$ is just the self energy divided by $p^2$
\BE
F(p^2/m_0^2)=\frac{\Sigma_L(p^2)}{\alpha p^2}
\EE
Its finite part is an explicit analytical expression that does not depend on any parameter and 
is evaluated in Ref.\cite{ptqcd2} by the sum of the finite parts of the graphs in Fig.~\ref{F1}.
The divergent part is canceled by the divergent part of $\delta Z$, yielding
a finite result. However, the finite part of $\delta Z$ depends on the subtraction point
of the renormalization scheme and 
its arbitrary choice should be compensated in Eq.(\ref{chi}) by a change of the finite multiplicative renormalization
factor $z$  that is arbitrary anyway.
Thus, the function  $F(s)$ is defined up to an additive (finite) 
renormalization constant.
Moreover, we can divide by $\alpha$ and absorb the coupling in the arbitrary factor $z$ yielding
\BE
z\>J(p)^{-1}=F(p^2/m_0^2)+F_0
\label{chi2}
\EE
where the new constant $F_0$ is the sum of all the constant terms.

As anticipated in the Introduction, the method is very predictive since the derivative of the inverse
dressing function is given by the derivative of the function $F$ and acquires a universal form that 
does not depend on any parameter and has been found in perfect agreement with the lattice data 
in the IR\cite{scaling}. Integrating back, the dressing function does depend on the integration constant $F_0$
which is related to the arbitrary choice of the subtraction point. The residual dependence of the propagator
on the choice of $F_0$ can be optimized in the Euclidean space by a comparison with the lattice data.
As shown in Refs.\cite{ptqcd2,scaling} an impressive agreement is obtained by taking $F_0=-1.05$ at the mass
scale $m_0=0.73$ GeV. Even if the optimal values might change as a function of temperature, we will take 
those values as fixed in the present paper. Their eventual variation would lead to a variational improvement
of the method at finite temperature.

\begin{figure}[b] 
\centering
\includegraphics[width=0.35\textwidth,angle=-90]{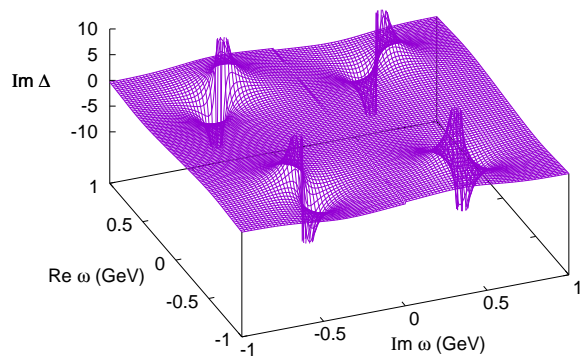}
\caption{Imaginary part of the one-loop gluon propagator $\Delta(p^2)$ evaluated by
Eq.(\ref{chi2}) for $F_0=-1.05$ and $m_0=0.73$ GeV in the complex plane $\omega=\sqrt{p^2}$. A very small
discontinuous structure can be seen on the real axis\cite{dispersion}.}
\label{F2}
\end{figure}

Denoting by $\omega$ the physical energy in Minkowski space and by ${\bf k}$ the three-momentum, so that
$p^2=\omega^2-{\bf k}^2$, we can study the propagator in the complex $\omega$-plane and in the long-wavelength limit
where $\omega=\sqrt{p^2}$. By a direct calculation through Eq.(\ref{chi2}), the dressing function has
two pairs of opposite complex conjugate poles at $\omega=\pm(m\pm i\gamma)$
where the real part $m=0.63$ GeV and the imaginary part $\gamma=0.48$ GeV.
A plot of the imaginary part of the gluon propagator $\Delta$ is shown in Fig.~\ref{F2}.

As discussed in Ref.\cite{dispersion}, the gluon propagator is very well approximated by the sum 
of the principal parts
\BE
\Delta_R(\omega)=\sum_{\pm} R_\pm\left[\frac{1}{\omega-(m\pm i\gamma)}-\frac{1}{\omega+(m\pm i\gamma)}\right]
\label{principal}
\EE
where $R_\pm$ are complex conjugate residues. The difference $\Delta-\Delta_R$ contains the very small
discontinue structure that can be observed on the real axis. That structure has no poles and is not
relevant for the asymptotic states, so that we can safely take $\Delta\approx\Delta_R$ in the following discussion.
A rational propagator like $\Delta_R$ in Eq.(\ref{principal}) was  
conjectured in the past and predicted by phenomenological models like the refined version\cite{dudal08,dudal08b,dudal11}
of the Gribov-Zwanziger model\cite{GZ}. As shown by Stingl\cite{stingl}, the existence of complex conjugate poles
would not violate unitarity or causality because the quasigluon would be canceled from the asymptotic states.
Actually, the existence of complex masses can be seen as a microscopic proof of confinement.
Following the argument of Stingl\cite{stingl} and taking $\Delta\approx\Delta_R$ we can Fourier transform
the propagator in Minkowski space and write
\BE
\Delta(t)=\int_{-\infty}^{\infty} \frac{{\rm d}{\omega}}{2\pi} \Delta_R(\omega) e^{i\omega t}
=-2 e^{-\gamma\vert t\vert}\>\vert R\vert \>\sin \left(m\vert t\vert+\phi\right)
\label{Dt}
\EE
where $R_\pm=\vert R \vert\> e^{\pm i\phi}$.
Even at $T=0$, the elementary excitations of the vacuum are short-lived quasigluons with an intrinsic
lifetime $\tau_0=1/\gamma$. In other words, all S-matrix elements involving one or more external gluons are
zero and cannot give rise to any unitarity or causality problem. During its short life, for $\vert t\vert\ll \tau_0$,
the quasigluon behaves like an eigenstate with energy $m$.

On the other hand, a thermal theory does not require the existence of asymptotic states and the quasigluons
contribute to the free energy and to other thermodynamic quantities. In that sense, they can appear as the
elementary degrees of freedom of a hot plasma above the deconfinement transition temperature $T_c$.
That motivates an extension to finite temperature of the massive expansion. 

We must mention that no complex poles were found by numerical solution of DSE in Minkowski space\cite{strauss}.
However, that calculation could be sensitive to the special ansatz that is used for the truncation of
the infinite set of integral equations. For instance, the existence of a peak on the real axis was claimed
in Ref.\cite{strauss} and replaced by a smooth function in Ref.\cite{glueballs} by the same authors.
Moreover, in any numerical calculation in the complex plane the choice of the correct Riemann sheet might not
be a simple task\cite{blaizot2005}. Thus, the present extension to finite temperature might also be  useful for 
establishing the genuine physical nature of the complex poles, since the results of standard perturbation 
theory should be approached when the temperature is high enough above $T_c$.

Another independent test for the reliability of the formalism would arise from a direct check 
of the gauge invariance of the poles, which is predicted by general arguments\cite{kobes90}
and formally shown in other schemes like the Gribov-Zwanziger framework\cite{capri}.
While the present work is in the Landau gauge, the formalism could be easily extended to a generic linear covariant gauge,
along the lines discussed in Ref.\cite{comitini}. When fixing the gauge by a covariant
term $(\partial A)^2/(2\alpha)$, with $\alpha\not=0$, the added mass $m_0$ in the quadratic part of the Lagrangian
should be replaced by a pure {\it transverse} mass-term, which is canceled by a {\it transverse} mass-counterterm $\delta \Gamma$ 
in the interaction.
Since the total Lagrangian is still unchanged, BRST invariance is not broken and the exact sum of the expansion
yields a vanishing longitudinal polarization. Thus, the dressed longitudinal propagator is known exactly and
is equal to the free one, $\Delta_L=-\alpha/p^2$, which is massless because no longitudinal mass was inserted in the quadratic part of
the Lagrangian.
The transverse propagator and its poles can be evaluated as before, by the massive expansion, using
the zeroth order propagator which has a massive transverse part, still given by Eq.(\ref{Dm}), and a massless (exact) longitudinal part.
However, since BRST is broken in the quadratic part of the Lagrangian, we do not expect that
the poles would be exactly invariant at any finite order of the approximation. Thus, their gauge dependence would measure the
accuracy of the approximation at any order.

\section{Pole trajectory at finite $T$}

The extension of the massive expansion to finite temperature is straightforward but tedious.
All graphs in Fig.~\ref{F1} acquire a thermal part and explicit expressions were reported in Ref.\cite{serreau}
for most of the required one-loop graphs. The new crossed graphs (including one
or more insertions of the counterterm $\delta \Gamma=m_0^2$) can be obtained by a simple derivative with
respect to $m^2_0$. 
All thermal parts are finite but require a numerical one-dimensional integration over the internal three-vector modulus
$q$. Explicit expressions will be published elsewhere.

The analytic continuation of integral functions is not trivial if singular points are integrated along the integration
path. As discussed in Ref.\cite{blaizot2005} we must check that the integration over $q$ on the real axis does not meet any
singular point of the logarithmic functions. Otherwise, a modified path must be chosen before the analytic continuation
can be undertaken. By inspection of the explicit expressions\cite{serreau}, branch cuts might be present, originating at the
singular branch point of the logarithmic function
\BE
L_\beta (z_\alpha)=\log\left[\frac{z_\alpha^2+\omega^2_{+,\beta}} {z_\alpha^2+\omega^2_{-,\beta}} \right]
\EE
where
$z_\alpha=i\omega\pm i\sqrt{q^2+\alpha^2}$ and $\omega^2_{\pm,\beta}=(q\pm k)^2+\beta^2$. Here $\alpha$ and $\beta$
are masses equal to $0$ or $m_0$,  while $k$ is the external three-vector modulus. Assuming the existence of a
branch point at $q=q_0$ on the real axis, it must satisfy 
\BE
\pm 2q_0k=\alpha^2-\beta^2-k^2+\omega^2\pm2\omega\sqrt{q_0^2+\alpha^2}
\label{zero}
\EE
where the $\pm$ signs are independent of each other. Taking $\omega=x+iy$ with $y>0$, the imaginary part of Eq.(\ref{zero})
gives $x=\mp \sqrt{q_0^2+\alpha^2}$ and substituting back in the real part we obtain $\omega^2_{\pm,\beta}+y^2=0$
which is never satisfied unless $y=\beta=0$. Thus if $\omega$ is not real, the branch point $q_0$ cannot be real and
the integral over $q$ on the real axis gives an analytic function of $\omega$. That condition is fulfilled around the
poles where $y\approx \gamma>0$. We can safely continue analytically the numerical thermal integrals from the Euclidean space ($x=0$, $y>0$) to the whole upper  half-plane. Moreover, in the large wavelength limit $k\to 0$, the logarithmic
function can be written as $L_\beta (z_\alpha)\approx \log\left[1+{\cal O} (k)\right]$ and the argument of the log never
vanishes. Thus, in that limit, there are no branch points at all and the numerical thermal integrals over $q$ can be safely continued to the whole complex $\omega$ plane.

The method could be used for a study of the full dispersion relations as functions of temperature and three-vector k, 
by following the location of the poles in the longitudinal and transverse projections of the propagator.
However, in the present paper we will content ourselves with the long wavelength limit $k\to 0$ where 
the longitudinal and transversal quasiparticles must have the same complex masses because there are no privileged directions. We checked that 
the poles of the longitudinal and transverse projections coalesce in that limit.

In principle, the additive renormalization constant $F_0$ and the mass scale $m_0$ should be optimized as functions
of the temperature. Here, they are fixed at their optimal value at $T=0$, neglecting their change at finite temperature.
Thus the method might be improved by some variational argument.

\begin{figure}[b] 
\centering
\includegraphics[width=0.35\textwidth,angle=-90]{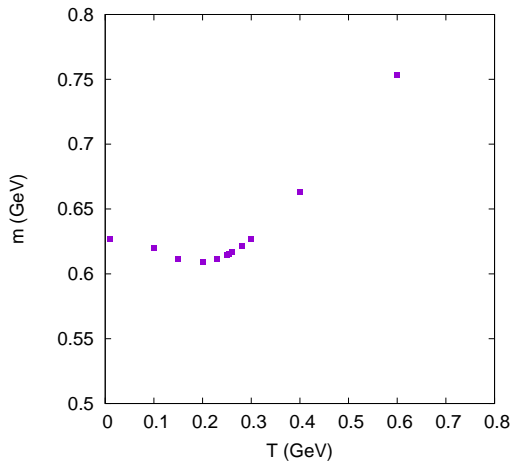}
\caption{The quasigluon mass $m$, i.e. the real part of the pole location in the complex $\omega$-plane for $k\to 0$,
is shown as a function of temperature for $F_0=-1.05$ and $m_0=0.73$ GeV.}
\label{F3}
\end{figure}

\begin{figure}[b] 
\centering
\includegraphics[width=0.35\textwidth,angle=-90]{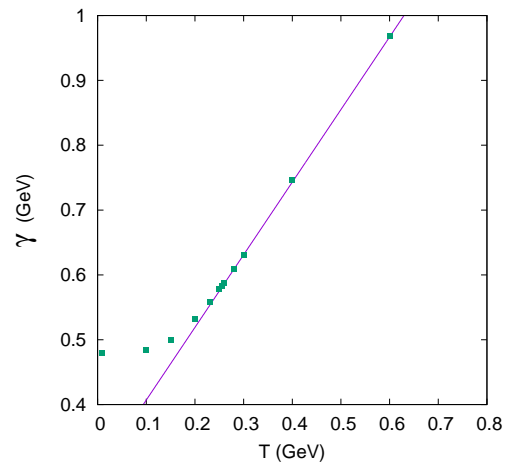}
\caption{The quasigluon damping rate $\gamma=1/\tau $, 
i.e. the imaginary part of the pole location in the complex $\omega$-plane for $k\to 0$,
is shown as a function of temperature for $F_0=-1.05$ and $m_0=0.73$ GeV. The straight line is the linear
function $\gamma=\gamma_0+b T$ with $\gamma_0=0.295$ GeV and $b=1.12$.}
\label{F4}
\end{figure}

In Fig.~\ref{F3} and Fig.~\ref{F4} the real part $m$ and the imaginary part $\gamma$ of the pole position in the complex
$\omega$-plane are displayed as functions of the temperature.

As shown in Fig.~\ref{F3}, the quasigluon mass $m$ is not monotonic. It decreases below $200$ MeV, reaches a minimum
at $T\approx 200$ MeV and then increases approaching a linear behavior above $400$ MeV. A non-monotonic behavior
was observed on the lattice for the longitudinal inverse propagator $1/\Delta(0)$ which 
defines a Debye mass\cite{silva,maas2012}. However the two
definitions of mass are quite different. While $\Delta(0)$ is a mass scale that depends on renormalization, 
gauge choice and polarization, the real part of the pole $m$ is the dynamical mass of the quasigluon, according to
Eq.(\ref{Dt}). It does not depend on the polarization  and is expected to be gauge invariant\cite{kobes90}. 
We checked that
the correct qualitative behavior is observed for the inverse propagator at $\omega=0$ in the long wavelength limit $k\to 0$,
where we find a non-monotonic longitudinal projection (Debye mass) and a 
monotonically increasing transverse projection (magnetic mass), as already shown in
Ref.\cite{serreau} by a massive expansion. We observe that different results are obtained reversing the order of the
two limits $\omega\to 0$ and $k\to 0$. For a finite $\omega$, the longitudinal and transverse projection must coincide
in the limit $k\to 0$, because there are no privileged directions. While for a finite $k$, the limit $\omega\to 0$ gives different definitions for the transverse and longitudinal propagators even when $k$ is very small, yielding different limits for the
Debye mass and the magnetic mass.

A more direct comparison can be made with phenomenological models that usually predict a non-monotonic quasigluon mass around the deconfinement transition\cite{werner2016,greco2011}. A minimum is found at the same temperature $T\approx 200$ MeV 
in Ref.\cite{werner2016}, while it is pushed above $1.5T_c$ in Ref.\cite{greco2011}. No discontinuity is found for the
mass in those models, even if that conclusion is in part the consequence of the details of the models that sometimes 
use a divergent ansatz for the mass from the beginning\cite{castorina2012}. A finite and continuous mass across the transition
has been explained in Ref.\cite{alba2012} by the coupling to the Polyakov loop. On that point, no reliable conclusion can
be reached by the present calculation since we expect that even a first-order transition might be rounded
by the one-loop approximation. However, Fig.~\ref{F3} shows a clear crossover: above $300$ MeV the mass approaches
the linear increasing behavior expected by perturbation theory\cite{braaten90}; decreasing the temperature below $200$ MeV the mass {\it increases} again because of the dynamical mass generation, leading to a strong deviation from the perturbative behavior,
with a residual intrinsic mass $m \approx 630$ MeV at $T=0$.

According to Eq.(\ref{Dt}) the imaginary part of the pole is the quasigluon damping rate $\tau^{-1}=\gamma$ and
is shown in Fig.~\ref{F4} as a function of temperature. Again we observe a remarkable crossover with a linear behavior above
$300$ MeV and a strong deviation from that behavior below $200$ MeV where the lifetime $\tau(T)$ saturates at the residual
intrinsic value $\tau_0^{-1}=\tau(0)^{-1}\approx 480$ MeV. As shown in the figure, the linear behavior is approached very quickly
above the transition and the quasigluon becomes an ordinary thermal quasiparticle with a damping rate that is very well approximated by the linear expression 
\BE
\tau^{-1}=\gamma_0+b\> T
\EE
where $\gamma_0=0.295$ GeV and $b=1.12$. Extrapolating to high temperature, we are tempted to compare the effective coupling
$b$ with the coefficient expected by perturbation theory $\gamma/T=a\>\alpha_s/2$, 
where the value $a\approx 6.6$ was evaluated by resummation of hard thermal loops\cite{braaten90}. 
The comparison would give a reasonable $\alpha_s=2b/a=0.34$ which is the actual coupling at $2$ GeV. 
We must mention that the standard perturbation theory fails to predict that coefficient unless the hard thermal loops
are resummed in a consistent way\cite{braaten90}. In the present massive expansion, the existence of a mass scale
$m_0$ inside the loops should mitigate the problem and the hard thermal loops are not expected to be relevant unless $T\gg m_0$.
Thus, in the range of temperature of Fig.~\ref{F4}, where $m_0=0.73$ GeV, the effect of hard thermal loops should be negligible.

Below $200$ MeV, where ordinary perturbation theory breaks down, the damping rate $\gamma$ deviates from the linear
behavior and saturates, because of the existence of a quasigluon intrinsic finite lifetime $\tau_0$ that does not arise
from thermal effects. The quasigluon remains short-lived even when $T\to 0$ and acquires a very different behavior than
ordinary thermal long-lasting quasiparticles.
Thus the crossover describes a transition from an intrinsically confined quasigluon to ordinary quasiparticle behavior.
It is remarkable that, albeit continuous, the transition takes place in the narrow range of temperature between
$200$ and $300$ MeV, that compares well with the critical temperature $T_c\approx 270$ MeV observed in the lattice\cite{lucini}.

\section{Discussion}
The massive expansion of Ref.\cite{ptqcd2} provides an analytical approach to QCD in the IR by perturbation theory.
Based on the original Faddeev-Popov gauge-fixed Yang-Mills Lagrangian in the Landau
gauge, the optimized expansion is in very good agreement with the lattice in the Euclidean space\cite{scaling} and 
allows for a straightforward analytic continuation to Minkowski space\cite{analyt}. Thus, the prediction of a finite
lifetime for the quasigluon can be seen as a microscopic proof of confinement.
At finite temperature, the pole trajectory describes a crossover from an intrinsically confined quasigluon for
$T< 200$ MeV, to an ordinary thermal quasiparticle for $T>300$ MeV. In the high-temperature phase the standard
linear behavior is recovered, strengthening the reliability of the method.

A physical description emerges, where the quasigluons are intrinsecally damped in the confined phase, with a
short lifetime $\tau_0$ that does not arise from thermal effects. Since the lifetime is even shorter in the
deconfined phase, with $\gamma (T)\approx m (T)$, one could even question what the word ``deconfinement'' really means. 
Moreover, even the usual notion of quasiparticle can be questioned when the distance of the pole from the real axis is so large 
that no relevant resonance structure can be observed in the imaginary part of the propagator on the real axis,
as shown in Fig.~\ref{F2}. However, even when $\gamma$ loses its meaning as a width of a broad resonance, according to
Eq.(\ref{Dt}) it retains its meaning as a damping rate $\tau^{-1}$ of the quasigluon short-lived intermediate state.

A parallel can be made with the scaling theory of localization and with the crossover from weak to strong disorder that
is observed in condensed matter. 
Assuming that in a disordered sample of size $L$ the electrons are
localized at the Fermi energy, any effect on the conductivity can only be observed if the
localization length $\xi$ of the states is shorter than $L$, while no phenomenological effect can be seen if
$\xi\gg L$ since the states appear as extended at the scale $L$.
At finite $T$, the electron coherence can only be probed at the dephasing scattering scale $L(T)\sim 1/T^n$ which is 
mainly due to inelastic scattering. Thus, at high temperature, when $L(T)<\xi$ the effects of disorder are weak and, 
even if the scattering length is shorter, the electrons can be described by ordinary perturbation theory because the 
intrinsic localization length $\xi$ is larger than the effective sample size $L(T)$. On the other hand,
in the low temperature limit,  $L(T)$ gets very large and when $L(T)>\xi$ the electrons appear as strongly localized,
with large deviations from the standard picture of thermal quasiparticles.

In heavy ion collisions, the time scale of the process is very large compared to the gluon lifetime $\tau$, so that the intermediate quasigluon states can only generate gluon-jet events. However, at the high temperature reached during the process,
the quasiparticles can only be probed during their very short thermal lifetime $\tau_{th}(T)\sim 1/T$ to be compared with
the intrinsic lifetime $\tau_0$ at $T=0$. Thus, in the high temperature limit, when $\tau(T)<\tau_0$, no phenomenological
evidence of confinement can appear in the thermodynamic behavior of the hot plasma. The quasigluon looks like deconfined
even if its lifetime is shorter. While in the low temperature limit, the thermal lifetime $\tau_{th}(T)$ gets very large,
so that when $\tau_{th} (T)>\tau_0$ the short intrinsic lifetime of the quasigluon emerges and the gluon looks
like confined, with large deviations from the predictions of standard perturbation theory.

Even if the crossover is found at the correct range of temperature, without adjusting any free parameter, the present
approach fails to predict the sharp first-order transition that is expected in pure SU(3) Yang-Mills  theory\cite{lucini}.
In principle, there is no evidence that the mass
and lifetime must be discontinuous at the transition. They have never been measured by lattice simulations and might 
not be the correct order parameter.
A continuous mass was found in Ref.\cite{alba2012} by assuming a coupling to the Polyakov loop. 
However, it is likely that any sharp change would be rounded by the present one-loop calculation.

It would be interesting to explore the effects of a further variational optimization of the mass parameter $m_0$ as
a function of temperature, along the lines that have been recently discussed in Ref.\cite{comitini}.
A temperature-dependent mass parameter was successfully employed in the phenomenological
approach of Ref.\cite{serreau} and could make some difference at the transition point.
However, while the mass term was added to the Lagrangian in that work and used as a fit parameter, here the mass would
arise from first principles by a variational approach to the exact Yang-Mills theory, yielding a very predictive
tool for the study of QCD in the IR.

\end{document}